\documentclass[aps,prb,twocolumn,showpacs,groupedaddress,superscriptaddress]{revtex4}
\usepackage{graphics,amsmath}
\usepackage{graphicx}
\usepackage{epsfig}

\def\co{\colon\negthinspace}

\newcommand{\la}{\langle}
\newcommand{\ra}{\rangle}

\newcommand{\om}{\omega}

\newcommand{\vf}{v_{\rm F}}

\begin{document}
\title{f-Sum Rule and Unconventional Spectral Weight Transfer in Graphene}

\author{J. Sabio}
\affiliation{Instituto de  Ciencia de Materiales de Madrid, CSIC,
 Cantoblanco E28049 Madrid, Spain}

\author{J. Nilsson}
\affiliation{Instituut-Lorentz, Universiteit Leiden,
P.O. Box 9506, 2300 RA Leiden, The Netherlands}

\author{A.~H. Castro Neto}
\affiliation{Department of Physics, Boston University, 590 
Commonwealth Avenue, Boston, MA 02215, USA}

\begin{abstract}
We derive and analyze the f-sum rule for a two-dimensional (2D) system of interacting electrons whose behavior is described by the Dirac equation. We apply the sum rule to analyze the spectral weight transfer in graphene within different approximations discussed in the literature. We find that the sum rule is generically dominated by inter-band transitions while other excitations produce sub-leading behavior. The f-sum rule provides strong constraints for theories of interacting electrons in graphene.
\end{abstract}

\date{May 2008}
\pacs{81.05.Uw,51.35.+a,71.10.-w} \maketitle
\section{Introduction.}

Exact sum rules provide a very insightful way to test the validity of certain approximations in the theory of the interacting electron liquid.\cite{PN66,GV05} In those systems, electrons interact via Coulomb forces, which are long-ranged, making perturbation theory questionable. A common way to deal with this issue is to perform an infinite diagrammatic resummation, such as the random phase approximation (RPA), that leads to the screening of the long range part of the interaction, effectively reducing it. The result can be used to obtain electron response functions, where the different excitations arising in the effective theory show particular features. This procedure is at the heart of the Landau Fermi liquid theory of He$^3$ and ordinary metals.\cite{GV05} In this context, the so-called f-sum rule provides a way to check if a particular resummation or approximation is describing the relevant excitations, and their relative importance.\cite{PN66}

The f-sum rule has been widely used in the context of 3D and 2D Landau-Fermi liquids, where interactions are switched on adiabatically from an initial gas of free Schr\"odinger electrons. In those systems, in the high density regime, the relevant excitations are correctly captured by the RPA resummation, leading to Landau quasiparticles and a collective plasmon mode. In the RPA approximation, perturbation theory can be cured once it is realized that the Coulomb potential is screened by the electrons. 

Recently, graphene has attracted a lot of attention due to the fact that its 
lattice geometry gives rise to Dirac-like electrons described by the 2D Dirac equation \cite{Graphene_rev}. This purely 2D crystal was isolated and characterized in the laboratory in 2005.\cite{Novoselov2} In contrast to an ordinary Fermi Liquid, neutral or undoped graphene is a semimetal, whose density of states vanishes at the Dirac point. In this way, the electrons have very poor screening properties, and the picture of quasiparticles composed of weakly interacting electrons could in principle fail. On the other hand, experimentally it is known that undoped graphene carriers behave as non-interacting Dirac electrons, so the role played by interactions in graphene is still unclear. Perturbative renormalization group (RG) arguments, which are equivalent to the RPA approximation, indicate that unscreened Coulomb interactions are marginally irrelevant, only leading to a logarithmic renormalization of the Fermi velocity.\cite{GGV94}  Nevertheless, this approach has been criticized recently from different directions. In Ref.~\onlinecite{Mischenko08}, it has been argued that new collective modes, arising from non-trivial vertex corrections beyond RPA, may play an important role in the low-energy description of graphene. A different result is obtained when interactions are studied in the tight-binding description of graphene. By using a ring diagram approximation,\cite{YT07} it has been showed that electrons in undoped graphene would behave essentially like a Fermi liquid, in contrast to the marginal Fermi liquid behaviour that arises from the RG calculation. Therefore, being yet a controversial issue, it is important to provide new tools to study the interacting problem in graphene, so as to reanalyze the relevance of the different electronic excitations in this system.  

In this paper we study the different approximations to electron interactions in doped and undoped graphene by means of the f-sum rule. In section II we review those approaches, putting special emphasis in the different excitations that are present. In section III we introduce the f-sum rule for 2D Dirac electrons and the constraints that imposes on approximations to the interacting problem. In section IV we study the spectral weight transfer in terms of the interaction strength. Section V contains the analysis of the results, while in section VI we summarize the main highlights of the work. We have included appendix \ref{sec:appendix} which contains a detailed derivation of the f-sum rule in different situations.

\section{The theory of interacting electrons in graphene.}

Due to the symmetry properties of the honeycomb lattice there are two inequivalent points in the Brillouin zone, K and K'. In the low-energy limit an expansion around those points gives an effective Dirac equation ruling the motion of the electrons. In dealing with long range interactions, the scattering is dominated by small momentum transfer allowing us to concentrate on each individual point and neglect inter-valley transitions. In this case, the Hamiltonian for the problem is given by (we use units such that $\hbar =1$):
\begin{equation}
H = v_F \sum_{\vec{k}} \Psi_{\vec{k}}^{\dagger} \left( \begin{array}{rr} -k_F & \phi_{\vec{k}}^*\\ \phi_{\vec{k}} & -k_F \end{array} \right) \Psi_{\vec{k}} + \sum_{\vec{q}} V_{\vec{q}} n_{\vec{q}}^{\dagger} n_{\vec{q}},
\label{HamiltonianDL}
\end{equation}
where $\phi_{\vec{k}} = k_x + i k_y$ and $\Psi_{\vec{k}}$ is a two component field operator, $v_F$ is the Fermi velocity, 
$k_F$ is the Fermi momentum (related to the chemical potential $\mu$ by $k_F = \mu / v_F $), $n^{\dagger}_{\vec{q}} = \sum_{\vec{k}} \Psi_{\vec{k}+\vec{q}}^{\dagger} \Psi_{\vec{k}}$ the density operator, and $V_{\vec{q}} = \frac{2\pi e^2}{q}$ the 2D unscreened Coulomb interaction. The non-interacting Hamiltonian can be easily diagonalized, giving rise to two different sub-bands or Dirac cones. In neutral graphene, the lower band is completely filled, while the upper one is empty. The single particle excitations in this case are inter-band excitations of electron-hole pairs. In the presence of a finite chemical potential (which can be tuned using a back gate in actual graphene devices) there are also intra-band electron-hole pairs in the same band. An important point to remark here is that electrons and holes in different sub-bands overlap, being responsible of some of the most striking phenomena observed in graphene, like the Klein Paradox.\cite{KG06}

In order to describe the different approximations to Hamiltonian Eq.~\eqref{HamiltonianDL}, we focus on the density-density response function, $\chi(\vec{q},\omega)$, which characterizes the response of the system to a density probe in linear response theory. Its imaginary part is related, by virtue of the fluctuation-dissipation theorem, to the dynamic form factor, $S_{\vec{q}}(\omega) = \sum_n | \la n | n_{\vec{q}}^{\dagger}|0 \ra|^2 \delta (\omega - \omega_{n0})$, that contains information on the relevant excitations $|n\ra$ connected to the ground state $|0\ra$ through the density operator ($\omega_{n0} = \omega_n - \omega_0$ is the energy of the excitation). 

{\it Non-interacting response function.} The density response of free Dirac electrons is well understood in both undoped and doped graphene. The former was studied first by Gonzalez {\it et al.}.\cite{GGV94} In this paper we are primarily interested in its imaginary part, which gives the possible dissipative processes in the system (it is also connected to the excitations in the system):
\begin{equation}
\Im \chi_0(\vec{q}, \omega) = - \frac{g_s g_v}{16}\frac{q^2}{\sqrt{\omega^2 - v_F^2 q^2 }} \Theta(\omega - v_F q)
\label{ImBareUndopedPol}
\end{equation} 
where $g_s = g_v = 2$ are respectively the spin and valley degeneracies. This function shows that free Dirac electrons have an infinite response at the threshold $\omega = v_F q$, and that possible dissipative processes are restricted to the region $\omega > v_F q$, corresponding to particle-hole continuum. 

The case of doped graphene is slightly more complicated. It was studied by Shung in relation to intercalated graphite,\cite{Shung86} and recently in Refs.~\onlinecite{WSSG07,HdS07} in relation to graphene. In the long-wavelength region, the imaginary part of the polarization function is given by:
\begin{equation}
\Im \chi_0^{doped} (\vec{q}, \omega) = \left\{ \begin{array}{rl} -\frac{g_s g_v}{16}\frac{q^2}{\omega} ,&  \omega > 2 E_F \\ 0, & v_F q < \omega < 2 E_F \\ - \frac{g_S g_v k_F}{2 \pi v_F} \frac{\omega}{\sqrt{v_F^2 q^2 - \omega^2}}, & \omega < v_F q \end{array} \right.
\label{ImBareDopedPol}
\end{equation}
where $E_F = v_F k_F$ is the Fermi Energy. This expression reflects the existence of two different particle-hole excitations with different responses: the intra-band ($\omega < v_F q$), and the inter-band ($\omega > 2 E_F$) excitations. In between there is a gap in the dissipative spectrum due to the fact that the displacement of the Fermi energy prohibits some inter-band transitions. Also in the doped case there is an infinite response at the threshold $\om = v_F q$.

{\it RPA theory of interacting electrons.} The RPA resummation of the bubble diagram, $\Pi(\vec{q},\omega) = - \chi_0 (\vec{q}, \omega)$  has been applied to a large variety of physical problems. As we mentioned in the introduction, when applied to the electron liquid it is known to capture the essential physics in the high density regime. Concerning undoped graphene, it was first employed by Gonzalez {\it et al.}\cite{GGV99} to argue for a marginal Fermi liquid behavior in this system. In the RPA approximation the imaginary part of the density-density response function is:
\begin{equation}
\Im \chi_{RPA} (\vec{q},\omega) = \frac{\Im \chi_0 (\vec{q},\omega)}
{[1 - V_{\vec{q}} \Re \chi_0 (\vec{q}, \omega)]^2 + V_{\vec{q}}^2 [\Im \chi_0 (\vec{q},\omega)]^2}.
\label{RPA}
\end{equation}  
Being proportional to the imaginary part of the bare susceptibility, the dissipative processes allowed in the RPA approximation are again restricted to the region over the threshold. In the particle-hole continuum the real part, $\Re \chi_0$, vanishes, leading to a simplified expression for the RPA susceptibility:
\begin{equation}
\Im \chi_{RPA} (\vec{q}, \omega) = - \frac{g_s g_v}{16}\frac{q^2 \sqrt{\omega^2 - v_F^2 q^2}}{\omega^2 - (v_F q)^2 + (\frac{\pi g}{2})^2 (\vf q)^2}.
\label{ImUndopedRPA}
\end{equation} 
We have defined the dimensionless coupling constant $g = (g_s g_v/4) e^2 / v_F$ (analogous to the fine-structure constant in quantum electrodynamics) that, in free standing graphene, is known to be of order $g \simeq 2$. This approximation is only valid for energies much smaller than the high energy cut-off of the theory, which is the energy band-width of the system $\Lambda_E$. Actually, in the whole regime where the imaginary part is known to be non-zero, so is also the real part by the Kramers-Kronig relation:

\begin{multline}
\Re \chi_0 (\vec{q}, \omega) = \frac{2}{\pi} P.V. \Bigl[ \int_0^{\Lambda_E} \frac{ d\omega' \omega'}{\omega'^2 - \omega^2} \Im \chi_0(\vec{q}, \omega') \Bigr] =
\\ = 
 \frac{g_s g_v}{16 \pi} \frac{q^2}{\sqrt{\omega^2 - (v_F q)^2}}
 \\ \times
 \log\Bigl[\frac{\sqrt{\Lambda_E^2-(v_F q)^2} + \sqrt{\omega^2-(v_F q)^2}}
 {\sqrt{\Lambda_E^2-(v_F q)^2} - \sqrt{\omega^2-(v_F q)^2}}\Bigr].
\label{ReBareUndopedPol2}
\end{multline}
In this work, in general we will use equation (\ref{RPA}) with non-zero real and imaginary parts. However, as we shall see in Section IV, including a non-zero real part is only relevant for energies very close to the cut-off. 

In contrast to the electron liquid, the poor screening properties of electrons in undoped graphene does not allow to recover the Fermi Liquid picture for Dirac electrons. Even though the screening cuts off the divergence at the threshold, the Coulomb potential remains long ranged and there are no plasmon modes. The relevant excitations in RPA are the inter-band electron-hole ones, just as in the bare theory. As we shall discuss later, however, this picture could be uncomplete, as it has been proposed recently that RPA does not capture the whole physics in undoped graphene,\cite{Mischenko08, KUC07} arguing that it is necessary to take into account also another sub-class of diagrams.

In doped graphene this is no longer a problem, and the RPA approximation has been shown to capture the essential physics.\cite{WSSG07,HdS07} This happens primarily due to the non-vanishing density of states at the Fermi level once the doping level is away from the Dirac Point. In the limit $v_F q << g E_F$, the imaginary part of the susceptibility is:
\begin{widetext}

\begin{equation}
\Im \chi_{RPA}^{doped} (\vec{q}, \omega) = \left\{ \begin{array}{rl} -\frac{g_S g_v}{16}\frac{q^2}{\omega},&  \omega > 2 E_F \\ - \frac{\omega_0^3}{4 e^2} \frac{q^{5/2}}{\omega^2}[1-\frac{\omega^2}{4 E_F^2}]\delta(\omega - \omega_0 q^{1/2}) , & v_F q < \omega < 2 E_F \\ - \frac{2 \omega}{\pi g_s g_v e^4 E_F}\sqrt{v_F^2 q^2 - \omega^2}, & \omega < v_F q \end{array} \right.
\label{ImDopedRPA}
\end{equation}

\end{widetext}
where $\omega_0 = (g_s g_v e^2 E_F/2)^{1/2}$ is the plasma frequency for graphene. In this case, the RPA shows the existence of a well-defined (undamped) plasmon mode in the region of prohibited inter-band excitations. The dispersion relation of this plasmon is given by $\omega = \omega_0 q^{1/2}$, which has the same q-dependence than the 2D electron gas (as opposed to the density dependence of the plasma-frequency, which is peculiar in the case of graphene). Besides, the intra-band and inter-band particle-hole excitations remain in the dissipative spectra. 

\begin{figure}
\begin{center}
\includegraphics[width=3in]{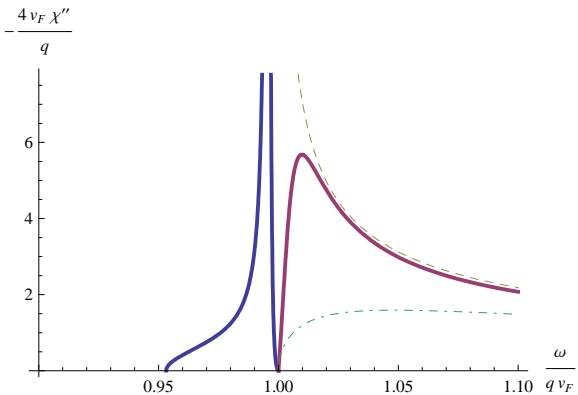}
\caption{Imaginary part of the susceptibility function for undoped graphene close to the resonance, for $g = 0.2$. The thick solid line shows the result of performing the resummation of loop and vertex diagrams. The dash-dotted line is the RPA approximation. The dashed line corresponds to the non interacting susceptibility. Notice that the former has a contribution in the region under the threshold, where Coulomb interaction gives rise to excitons and a damped collective mode (the highest peak).}
\label{Polarizability}
\end{center}
\end{figure}

{\it Beyond the RPA theory in undoped graphene.} As we have said, the validity of the RPA approximation in undoped graphene has been questioned. \cite{Mischenko08, KUC07} The idea is that close to the resonance ($\omega = v_F q$), there are non-RPA contributions which can only be neglected in the limit of infinite number $N_f$ of fermion species (for graphene, we have $N_f = g_s g_v = 4$). When the condition $N_f \simeq \log(v_F q / |v_F q - \omega|)$ is fulfilled, it was shown that the correct susceptibility is given by:
\begin{equation}
{\cal \chi}(q,\omega) = \frac{\chi_V(q,\omega)}{1 - v_q \chi_V (q,\omega)}
\label{resum}
\end{equation} 
where $\chi_V$ is the ladder susceptibility:
\begin{equation}
\chi_V (q,\omega) = - \frac{q}{v_F g} \frac{1 + \frac{2}{\pi} \arcsin(x) - (1 + \frac{2}{\pi}x)\sqrt{(1-x^2)}}{\log(\frac{v_F q}{|v_F q - \omega|}) x \sqrt{(1 - x^2)}}
\label{ladderPol}
\end{equation}
having defined $x=\frac{g}{2\sqrt{2}} \sqrt{\frac{v_F q}{v_F q -\omega}} \ln  (\frac{v_F q}{|v_F q -\omega|})$. The imaginary part of this function is shown in Fig.~\ref{Polarizability}, together with the other susceptibilities analyzed in the text for undoped graphene. The most striking feature of this susceptibility is the new threshold of excitations that appears under $\omega = v_F q$. Due to the Coulomb interaction, electron-hole pairs reduce their energy, giving rise to excitons. The new lower threshold for excitations is hence no longer $\omega = v_F q$, but a new one determined through the condition $x = 1$. Also, the susceptibility shows a peak in this region, where the real part of the susceptibility becomes positive, which can be related to a damped collective mode. As the interaction strength is increased, the threshold is pulled downwards, while the collective mode becomes less damped. In the limit $g \gg 1$ the latter is no longer damped, and its dispersion relation can be found analytically: $\omega_q = v_F q (1 - e^{-N_f})$. In this region the contribution from excitons is negligible. Notice, however, that for intermediate coupling the threshold goes beyond the region of applicability of this susceptibility, which is only valid close to the resonance $\omega = v_F q$. For those values of the parameters a more detailed study should be performed.

\section{The f-sum rule for 2D Dirac electrons}

As mentioned in the introduction, exact sum rules are very useful as a check of different approximation schemes in interacting electron systems. In particular, it is known that in the electron liquid theory any approximation to the interacting Hamiltonian has to fulfill the identity:
\begin{equation}
  - \int_0^{\infty} \frac{d\omega}{\pi} \, \omega \, \Im \chi(\vec{q},\omega) = \frac{q^2 N}{2 m}
\label{fsumruleSchrodinger}
\end{equation}
where $N$ is the electron density (with volume normalized to one). This identity, called the f-sum rule after the oscillator parameters $f$ in atomic physics, is justified for Schr\"odinger electrons, but no longer applies to the case of Dirac. The corresponding f-sum rule for Dirac electrons can be derived following the steps of Nozi\'{e}res and Pines.\cite{PN66} Details of the derivation are given in the App.~\ref{sec:appendix}. The result is:
\begin{equation}
\int_0^{\Lambda_E} d\omega \, \omega \, \Im \chi(\vec{q},\omega) = - \frac{g_S g_v q^2 \Lambda_E}{16} ,
\label{grapheneSumRule}
\end{equation}
where $\Lambda_E$ is the energy cut-off (i.e. of the order of the band-width). The f-sum rule was studied earlier in the context of the true 3D relativistic Dirac problem \cite{GD82} with infinite band-width (no lattice). In that case, the sum rule gives an infinite contribution that is considered part of the ``vacuum'' energy and hence unmeasurable. In a system with a finite band-width, which is the case of graphene, such a cut-off dependent contribution represents a real response of the system (the inter-band electron-hole excitations), and hence is non-negligible. This issue was raised originally in Ref. \cite{Cenni01}, where a relativistic generalization of the f-sum rule is discussed too.  Notice that, as shown in the appendix, in order to deduce a non-zero f-sum rule for Dirac electrons we must proceed carefully, since conventional general definitions of the f-sum rule do not apply immediately due to the peculiar structure of the Dirac Hamiltonian.   

The f-sum rule for the imaginary part of the susceptibility can be related to a f-sum rule for the optical conductivity straight-forwardly. In a recent work \cite{GSC07}, the latter has been studied for the whole band of graphene, by using a tight-binding description. Though this approach is more realistic when applied to graphene, the f-sum rule for the continuous case provides more insights when studying models of interacting electrons in the low-energy regime, which is our interest in the present work.  

The f-sum rule for Dirac electrons have many peculiarities that are not present in the usual Fermi liquid. The contribution from the lower filled band makes the sum rule cut-off dependent. This is easy to understand as there are inter-band particle-hole transitions for arbitrary energy if we do not cut off the spectrum. Moreover, it does not depend on the level of doping (or the chemical potential) since  Eq.~\eqref{grapheneSumRule} is valid also for doped graphene. Even more striking, the sum rule applies too in the case of massive Dirac fermions (i.e. in a gapped system) as long as the gap is much smaller than the cut-off: $m \ll \Lambda_{E}$. A gap can be generated when a substrate breaks the graphene sublattice symmetry.\cite{Lanzara08}

The f-sum rule is a statement on the conservation of particle number. As interactions are switched on adiabatically, the number of particles does not change, and the sum rule must be fulfilled by any susceptibility calculated approximately from the interacting Hamiltonian. The effect of interactions is usually a rearrangement of the contribution to the f-sum rule by the different excitations. This is immediately related to the spectral weight $(|\la n|n_{\vec{q}}^{\dagger}|0 \ra|^2)_{ex}$ of the excitation, a quantity that is relevant in dissipative processes,\cite{PN66} and other physical properties such as van der Waals forces between neutral systems.\cite{Anderson83}

\section{Transfer of spectral weight}

We now turn to the application of the exact f-sum rule, derived in the last section, to particular approximations for the Coulomb-interacting Dirac problem, stressing the (re)distribution of the spectral weight among different excitations. Firstly, we study the case of free electrons, in both doped and undoped graphene. Then we discuss the more controversial issue of graphene with interactions. Here, the discussion of Section II on the different approximations to the problem will be especially relevant. From here on we concentrate on the particular case of $N_f = 4$ fermion species, which is the case of graphene.

\subsection{Free Dirac electrons}

If graphene is undoped, i.e. the Fermi level is located at the Dirac Point, the susceptibility is given by Eq.~\eqref{ImBareUndopedPol}. This expression saturates the f-sum rule for Dirac electrons, as expected, meaning that the contribution from inter-band particle-hole excitations has a $q^2$ dependence and that the number of those excitations is only limited by the high energy cut-off. 

When the level of doping is changed away from half filling, we have two different particle-hole excitations: intra-band and inter-band. As the f-sum rule does not depend on doping, the spectral weight must be distributed among those. By using Eq.~\eqref{ImBareDopedPol}, in the long-wavelength limit, we find a contribution from inter-band transitions:
\begin{equation}
\frac{q^2}{4}\int_{2E_F}^{\Lambda_E} d\omega = \frac{q^2 \Lambda_E}{4} - \frac{q^2 E_F}{2}
.\label{InterbandDoped}
\end{equation}
On the other hand, the contribution from the intra-band excitations is:
\begin{equation}
\frac{2 k_F}{v_F \pi} \int_0^{v_F q} d\omega \frac{\omega^2}{\sqrt{v_F^2 q^2 - \omega^2}} = \frac{q^2 E_F}{2} ,
\label{IntrabandDoped}
\end{equation}
meaning that there is an exact transfer of spectral weight between the two types of excitations that is proportional to the level of doping. When the Fermi level is changed, some inter-band transitions are prohibited by Pauli exclusion, and they no longer saturate the f-sum rule. Both excitations have a contribution proportional to $q^2$. However, the contribution from intra-band excitations is not cut-off dependent, but density (or doping) dependent. As we shall see when connecting the interactions, those excitations resemble those of the conventional Fermi liquid, where the contribution to the f-sum rule due to particle-hole excitations does depend on the density of electrons.

\subsection{Interacting Dirac electrons}

{\it RPA in undoped graphene.} As we have seen in Section II, the simplest approximation we can employ to study the role of interactions is the RPA. For an undoped sheet, we use equations (\ref{RPA}) and (\ref{ReBareUndopedPol2}) and find that the RPA approximation fulfills the f-sum rule. The main issue related to the RPA comes when analyzing the spectral weight. In principle, in undoped graphene at the RPA level, there are no new low-energy excitations besides the particle-hole. However, it can be seen that the low-energy sector loses some spectral weight. If we take an intermediate energy scale, say $\Lambda_I$, such that $v_F q \ll \Lambda_I \ll \Lambda_E$ and integrate over energy in this range, we get:
\begin{equation}
\int_0^{\Lambda_I} d \omega \omega \Im \chi_{RPA} (\vec{q}, \omega) = - \frac{q^2 \Lambda_I}{4} + \frac{\pi^2}{16} g v_F q^3.
\end{equation}
Thus there is a cut-off independent loss of spectral weight in the low-energy sector. This is an unconventional result in the sense that the usual RPA approximation for Fermi liquids only rearranges the low-energy spectral weight, and does not couple to the high energy sector of the theory. On the contrary, in undoped graphene it seems to be a spectral transfer to the high energy sector. In particular, the spectral weight is distributed close to the cut-off energy scale, where we find a new resonance and even a plasmon condition that is fulfilled. Of course, the particular kind of high energy excitations cannot be described by our effective theory, which only applies to the low-energy regime. The transfer is proportional to the interaction strength, $g$, and it depends on the momentum as  $q^3$, instead of the typical $q^2$ dependence of single particle excitations, being negligible at leading order in the long-wavelength limit.

{\it RPA in doped graphene.} In the doped case, the RPA susceptibility is given by Eq.~\eqref{ImDopedRPA}, which predicts three relevant excitations: inter-band and intra-band particle-hole, and a plasmon mode. The contribution from the inter-band excitations is again given by Eq.~\eqref{InterbandDoped}, which states that these excitations have lost a contribution to the f-sum rule given by $E_F q^2 /2$. As opposed to the non-interacting theory, this contribution is not transferred to the intra-band particle-hole transitions. In this case their contribution is:
\begin{equation}
\frac{1}{2 \pi e^4 E_F} \int_0^{v_F q} d\omega \omega^2 \sqrt{v_F^2 q^2 - \omega^2} = \frac{v_F^2 q^4}{32  g^2 E_F},
\end{equation}
which is proportional to $q^4$ and thus much smaller in the long-wavelength limit than the loss from the inter-band transitions. As in the case of the Fermi liquid, it is the plasmon mode which absorbs most of the spectral weight:
\begin{eqnarray}
\frac{(2 e^2 E_F)^{3/2} q^{5/2}}{4 e^2}\int_{v_F q}^{2E_F} \frac{d\omega} {\omega}[1-\frac{\omega^2}{4 E_F^2}]\delta(\omega - \omega_0 q^{1/2}) = 
\nonumber
 \\
=  \frac{q^2 E_F}{2}\,\,.
\end{eqnarray}
Hence, in the asymptotic limit of $q \to 0$, the spectral weight of the intra-band excitations is exactly absorbed by the plasmon. 

In the interacting system the response is dominated by the plasmon, instead of the intra-band particle - hole excitations.  Notice, however, that graphene, due to its particular two-band structure, is different from the electron gas in the sense that the main contribution comes always from the inter-band transitions, which essentially saturates the f-sum rule. Despite this fact, as will be shown in next section, the existence of this collective mode can have measurable consequences.

Far above the energy scale $E_F$, we recover the same polarizability that for undoped graphene, as expected as now the only relevant excitations are the intra-band. Hence, a transfer of spectral weight to the high energy sector is also observed. Notice that, at the same order of this transfer, $q^3$, further rearrangements of spectral weight occur in the low-energy sector, as can be shown by computing next-order corrections to expression (\ref{ImDopedRPA}).  

{\it Beyond RPA in undoped graphene.} As mentioned above, in doped graphene the RPA approximation seems to work well, so we concentrate now in undoped graphene. As we have seen, when vertex corrections are taken into account, the polarization function close to the threshold $\omega = v_F q$ is given by Eqs.~\eqref{resum} and \eqref{ladderPol}. As this approximation does not describe the whole energy range, we cannot verify immediately the validity of the f-sum rule. However, we know that for high energies the RPA approximation should be valid, so we expect a loss of spectral weight similar to the one found before. The most relevant question is how the spectral weight is distributed among the new excitations that arise.

We would like to study the problem of spectral weight transfer when vertex corrections are added. As its contribution to the f-sum rule is related to the role they play in the response of the system, we could expect that new excitations can give deviations from the free electrons picture. However, the new modes found in Ref. \onlinecite{Mischenko08} can only be calculated reliably close to $\omega = v_F q$, and one cannot integrate the susceptibility given in Eq. (\ref{ladderPol}) in the entire energy range. Nevertheless, we can concentrate on the transfer of spectral weight close to $\omega = v_F q$ by defining the following quantity:
\begin{eqnarray}
S_p(g,C) = \frac{4}{v_F q^3} \int_0^{\Lambda_q} d\omega \, \omega \, \Im \chi(q, \omega) \, ,
\end{eqnarray}
where $\Lambda_q = C v_F q$ and $C$ is a constant that has to be chosen appropriately. If we define $s = \omega / v_F q$ we find:
\begin{eqnarray}
S_p(g,C) = \int_0^{C} ds s (4v_F \Im \chi(q, s)/q)
\end{eqnarray}
which is momentum independent but cut-off and interaction strength dependent. We find that the approximation of Eq. (\ref{ladderPol}) works well for $C - 1 \sim 10^{-1}$ (in such a way that $N_f = 4 \simeq \log(v_F q/(\Lambda_q-v_F q))$) and this is the value we use from now on. Readily, we notice that all the contributions to the f-sum rule that come from this region are of order $q^3$, and hence all the spectral rearrangement because of interactions will be of subleading order.

\begin{figure}
\centering
\includegraphics[scale=.35]{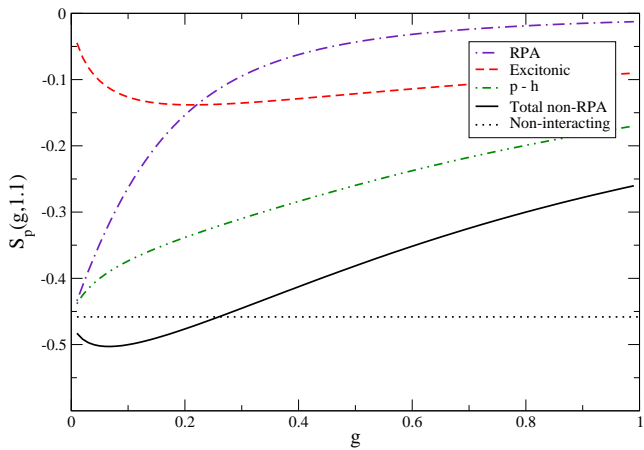}
\caption{Contributions to the f-sum rule in the low-energy region for undoped graphene. The integrals are done with a high energy cut-off such that $\omega < 1.1 v_F q$. The dotted line corresponds to the non-interacting theory. The dash-dotted line shows the RPA contribution. The solid line is the non-RPA contribution, sum of the excitonic contribution (under the threshold, $\omega < v_F q$) and the particle-hole contribution (over the threshold, $\omega > v_F q$). }
\label{SpectralTransfer}
\end{figure}

We analyze in the following the dependence on the interaction strength $g$. In Fig.~\ref{SpectralTransfer} we plot the various contributions to $S_p(g,1.1)$. This Figure shows a similar behavior for the electron-hole background in both RPA and non-RPA approximations: as the interaction strength is increased, the spectral weight covered in the integration is smaller. Notice that this should not be related only to the transfer of spectral weight to the high-energy sector (which is linear in $g$), but mostly to the fact that the maximum of the imaginary part of the susceptibility is shifted away from the threshold as the interaction grows. More interesting is the behavior of the excitonic part: it shows a maximum for $g \simeq 0.2$, and then starts to decrease.  

In order to understand this behavior, we notice that the contribution to the f-sum rule from the plasmon can be easily worked out in the limit $g \gg 1$, where the plasmon gets well defined:
\begin{equation}
\int_{pl} d\omega \omega (\Im \chi)_{pl} = - \frac{v_F q^3}{4}[\frac{8 e^{-N_f}}{g}(1 - e^{-N_f})] ,
\end{equation}
and where we have used the analytical plasmon dispersion relation for generic number of fermion species (remember that in the case of graphene we have $N_f = 4$), $\omega(q) = v_F q (1 - e^{-N_f})$, and its spectral weight, $|\la n|n_q^{\dagger}|0 \ra |^2_{pl} = 2 v_F q^2 / g e^{-N_f}$. We see that the plasmon contribution to the f-sum rule decreases as the interaction strength grows. On the other hand, although the excitonic domain gets broader (in thus out of the region of validity of our approximation), its contribution is also negligible for large interactions. Hence, the appearance of a maximum in the contribution to the f-sum rule under the threshold can be understood as an interplay between excitonic response and plasmon response.

Anyway, our results show that there is no important transfer of spectral weight from the particle - hole excitations to the new modes predicted in the context of vertex corrections, as they are of order $q^3$. As in principle this approximation only rearranges spectral weight close to the threshold, we expect a flow of spectral weight to the high energy sector of the theory as the interaction strength is increased, as occurred in the RPA approximation.

\section{Analysis of the results}

The different contributions to the f-sum rule have been summarized in Table \ref{table:contributions}. The Dirac liquid shows particular features from the point of view of spectral weight transfer and contributions to the f-sum rule. The results provide many insights into the role played by Coulomb interactions in graphene.

The f-sum rule is essentially saturated by the particle-hole excitations in the long-wavelenght limit. This is true for small doping levels compared to the cut-off, as the intra-band excitations have the largest contribution. In the case of doped graphene, Coulomb interactions give rise to a collective mode which absorbs to leading order all the spectral weight of inter-band excitations. However, this does not remain true in the case of undoped graphene, where all the spectral (re)arrangement due to the Coulomb interaction is always at subleading order, $q^3$. New excitations which could be arising close to the threshold contribute much less to the f-sum rule than the particle - hole ones, even when the interactions strength is infinitely large. This effect must be understood not only because of the subleading dependence of those excitations, but because of the spectral weight transfer to the high energy sector we have found in every approximation studied for the interacting Dirac liquid. Hence, the latter should be a remarkable feature of this system. 

An apparently related behavior has been described in high-Tc literature, where spectral weight transfer between different energy scales has been reported in several works. \cite{Hirsch92, Millis03} In the context of strongly correlated systems, a spectral weight transfer from the high energy sector to the low-energy one has been related to possible issues arising in the definition of the low-energy theory. \cite{EMS93} In this sense, studies concerning the whole graphene band-structure could help to clarify this issue. \cite{YT07}

When turning to graphene, the Dirac liquid is only a low-energy approximation to the electronic structure, and the f-sum rule derived here does not cover the whole band. However, from our results we expect in this case a transfer of spectral weight beyond the artificial cut-off that we introduced to delimitate the continuous description. Actually, some related effect has been observed in the Coulomb impurity problem in graphene,\cite{PNC07} where a bound state appears beyond the band, i.e., in the high energy sector. 

The relative lack of importance of Coulomb effects in graphene is in agreement with the experimental observation that electronic carriers are very well described by free Dirac fermions. This is specially true for undoped graphene, but also applies to doped graphene, where dissipative processes would be dominated by the excitation of the incoherent particle - hole background. Systematic experimental studies, however, should give a clear trace of the plasmon mode in doped graphene, which actually seems to have been observed in ARPES experiments \cite{Rotenberg07, Polini08, HdS08} and could be in principle detected in inelastic x-ray scattering experiments, where information about the dynamical structure factor can be extracted. Another physical quantity where plasmons could be important is the van der Waals force between graphene and other neutral systems, i.e., another graphene layer or a substrate \cite{sabio08}. In perturbation theory, the first correction to the energy of the system can be shown to be:
\begin{equation}
E^{(2)} = - \sum_q \int d\omega V_q^2(z)\chi_G(\vec{q},i\omega) \chi_G(\vec{q},i\omega)
\label{VdW}
\end{equation}
In undoped graphene, despite the new excitations predicted, this correction can be shown to be $E^{(2)} = - V_i(\frac{z_0}{z})^3$ in leading order,\cite{DWR06} where $z_0$ and $V_i$ are, respectively, typical distance and energy scales of the problem, and $z$ is the distance between the two graphene sheets. On the contrary,  in doped graphene, the response of the plasmon gives a new contribution:
\begin{equation}
   E^{(2)} = - V_i(\frac{z_0}{z})^3 - V_{pl} (\frac{z_0}{z})^{5/2}
\label{VdWCont}
\end{equation}
For an intermediate regime of distances, the dominant interaction is the one coming from the response of particle - hole excitations, as $V_i \gg V_{pl}$. The leading contribution, however, should be the one coming from the plasmon, as it is the one which decays more slowly.

\begin{widetext}
\begin{table}
\begin{tabular}{||l||c|c|c|c|c||}
\hline \hline & Free  & Free & RPA & RPA  & Beyond RPA \\
& Undoped & Doped & Undoped & Doped & Undoped\\
\hline
Inter-band  & $\Lambda_E q^2$  & $ (\Lambda_E - 2 E_F)q^2$ &$ \Lambda_E q^2$ & $(\Lambda_E - 2 E_F)q^2$ & $\Lambda_E q^2$  \\
Intra-band & -  & $2 E_F q^2$ & - &$v_F^2 q^4 / g^2 E_F$ & - \\
Plasmon/excitons  & -  & - & -&  $2 E_F q^2$ & $v_F q^3/g$  \\
High energy sector& -  & - & $g v_F q^3$ & $g v_F q^3$ & $g v_F q^3$  \\ 
 \hline \hline
\end{tabular}
\caption{Contributions to the f-sum rule from the different excitations present in the theories of non-interacting and interacting electrons in graphene studied in this work, both for doped and undoped graphene.}
\label{table:contributions}
\end{table}
\end{widetext}

\section{Conclusions}

We have derived the f-sum rule for Dirac electrons and applied it to study the spectral transfer in different approximations for interacting electrons in graphene. The f-sum rule, being an statement of the particle number conservation, is not interaction dependent and hence has to be fulfilled by any theory of the interacting Dirac liquid. Hence, it provides an insightful tool to study the effects of electron-electron interactions in the low-energy regime of graphene samples. 

We have shown that the RPA theory for undoped and doped graphene fulfills this identity, even though we find that its behavior turns out to be quite unconventional. Instead of having a spectral weight transfer among states around the same energy scale, the transfer involves two different scales, with one being on the order of the cut-off. When vertex corrections are included, this behavior is not changed, being a feature of the Dirac liquid. Despite the f-sum rule derived here only applies to low-energy graphene, we expect a similar transfer of spectral weight to occur once the whole band is taken into account.  

Besides, we have studied the relative importance of the different excitations predicted close to the Dirac point in graphene. The f-sum rule is essentially saturated by the inter-band particle-hole excitations, though in doped graphene the collective plasmon acquires some importance which could have measurable consequences in ARPES or x-ray inelastic scattering experiments. This is not the case in undoped graphene, where any new excitation coming from Coulomb interaction is essentially negligible compared to the electron-hole ones, as far as the spectral weight is concerned. Surprisingly, this feature remains true even for the collective mode predicted when vertex corrections are taken into account. Hence, its possible experimental observation, as well as the observation of other excitations different from the electron-hole ones in undoped graphene, can be technically challenging. As far as the measurements imply a certain average of the spectral weight, we expect that undoped graphene will respond to experimental probes essentially as a non-interacting system.  

\acknowledgments

We acknowledge F. Guinea for useful insights and B. Valenzuela for carefully reading the manuscript and for stimulating discussions. J.N. was supported by the Dutch Science Foundation FOM.  J.S. was supported by MEC (Spain) through
grant FIS2007-65723 and the Comunidad de Madrid,
through the program CITECNOMIK, CM2006-S-0505-ESP-0337. J.S. also wants to acknowledge the I3P Program from the CSIC for funding.

\appendix

\section{Derivation of the f-sum rule}
\label{sec:appendix}

The f-sum rule can be derived by relating the integral over the first momentum of the structure function and the following commutator:\cite{PN66}
\begin{equation}
\la 0|[[n_{\vec{q}}, H],n_{\vec{q}}^{\dagger}]|0 \ra = 2 \int_0^{\infty} d\omega \omega S_{\vec{q}}(\omega).
\end{equation} 
For what concerns us, it is more useful to express this identity in terms of the density-density correlation function, $\chi(\vec{q},\omega)$:
\begin{equation}
\la 0|[[n_{\vec{q}}, H],n_{\vec{q}}^{\dagger}]|0 \ra  = - \frac{2}{\pi} \int_0^{\infty} d\omega \omega \Im \chi(\vec{q},\omega).
\label{sumrule}
\end{equation}

\subsection{Undoped graphene}

In graphene, at low energies, if we focus on a single valley and neglect the spin degree of freedom, the Hamiltonian is:
\begin{equation}
H_0 = v_F \sum_{\vec{k}} \Psi_{\vec{k}}^{\dagger} \left( \begin{array}{rr} 0 & k_x - i k_y\\ k_x + i k_y & 0 \end{array} \right) \Psi_{\vec{k}} ,
\label{Hamiltonian}
\end{equation}
where $\Psi_{\vec{k}}^{\dagger} = (a_{\vec{k}}, b_{\vec{k}})$ is a spinor, being $a$ and $b$ operators referring to the two different sublattices in the unit cell of the honeycomb lattice. 
In the same basis, the density operator is given by:
\begin{equation}
n^{\dagger}_{\vec{q}} = \sum_{\vec{k}} \Psi_{\vec{k}+\vec{q}}^{\dagger} \Psi_{\vec{k}} = \sum_{\vec{k}} (a_{\vec{k}+\vec{q}}^{\dagger} a_{\vec{k}} + b_{\vec{k}+\vec{q}}^{\dagger} b_{\vec{k}}).
\end{equation}  
The Coulomb Interaction has the usual form $H_{int} = \sum_{\vec{q}} V_{\vec{q}} n_{\vec{q}}^{\dagger} n_{\vec{q}}$ (up to a term that is proportional to the number of particles in the system that is not relevant for this discussion), so it is simple to show that the condition $[n_q, H_{int}] = 0$ is realized. So wee see clearly the advantage of the f-sum rule: it can be calculated within the non-interacting theory, but it must be satisfied also by the interacting one.

\begin{figure}[htb]
\centering
\includegraphics[scale=1]{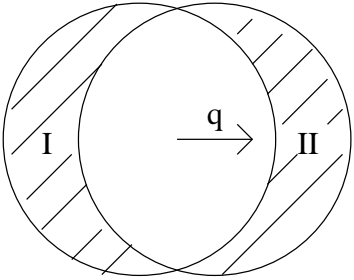}
\caption{Regions subtracted in the calculation of the f-sum rule}
\label{figure_intregions}
\end{figure}

With these expressions, the commutators in Eq.~\eqref{sumrule} can be evaluated. We notice that the first commutator is nothing but the particle conservation equation written in the momentum space:
\begin{equation}
[n_{\vec{q}}, H] = \vec{q} \cdot \vec{J}_{\vec{q}},
\end{equation}
where $\vec{J}_{\vec{q}} = v_F \sum_{\vec{k}} \Psi_{\vec{k}}^{\dagger} \vec{\sigma}  \Psi_{\vec{k}+\vec{q}}$ is the velocity (current) operator. So the double commutator reads:
\begin{equation}
[[n_{\vec{q}},H],n_{\vec{q}}^{\dagger}] = -v_F \sum_{\vec{k}} (\Psi_{\vec{k}+\vec{q}}^{\dagger} \vec{q} \cdot \vec{\sigma} \Psi_{\vec{k}+\vec{q}} - \Psi_{\vec{k}}^{\dagger} \vec{q} \cdot \vec{\sigma} \Psi_{\vec{k}}).
\end{equation}
Now we should proceed carefully with this result. In principle, our free theory describes massless electrons with an unbounded linear dispersion relation. But when handling operators that are defined in an unbounded region, we are not allowed, for instance, to simply state that $\sum_k G(k+q) = \sum_k G(k)$. The same issue can be found in the theory of the one dimensional electron liquid when calculating the commutator $[n_q, n_q^{\dagger}]$, which turns out to be nonzero, giving rise to the anomalous commutator problem.\cite{GV05} Once we work with unbounded operators, we need to refer the calculation to bounded quantities, which usually are defined with respect to the ground state value. Then, we define normal ordered operators:
\begin{equation}
\co G(k) \co = G(k) - \la 0|G(k)|0 \ra .
\end{equation}
As now the normal ordered operator is bounded, it satisfies $\sum_k \co G(k+q) \co = \sum_k \co G(k) \co$. Applying this rule to our commutator, we get:
\begin{multline}
[[n_{\vec{q}},H], n_{\vec{q}}^{\dagger}] =
\\ =
 -v_F \sum_{\vec{k}} \bigl( \la 0|\Psi_{\vec{k}+\vec{q}}^{\dagger} \vec{q} \cdot \vec{\sigma} \Psi_{\vec{k}+\vec{q}}|0 \ra - \la 0|\Psi_{\vec{k}}^{\dagger} \vec{q} \cdot \vec{\sigma} \Psi_{\vec{k}}|0 \ra \bigr),
\end{multline}
where the ground state for undoped graphene consists on the lower band completely filled (Dirac sea) and the upper band completely empty. 
The result involves the difference between two infinite sums, something not defined a priori. To compute it, we need to regularize somehow the sums, for instance, by using an ultraviolet cut-off. Besides, we need to switch to the diagonal basis. As usual, this is done with the unitary transformation that diagonalizes Hamiltonian (\ref{Hamiltonian}):
\begin{equation}
U_{\vec{k}} = \frac{1}{\sqrt{2}}
\begin{pmatrix}  e^{-i\frac{\theta_{\vec{k}}}{2}} & e^{-i\frac{\theta_{\vec{k}}}{2}} \\  e^{i\frac{\theta_{\vec{k}}}{2}} & - e^{i\frac{\theta_{\vec{k}}}{2}} \end{pmatrix},   
\end{equation} 
which leads to the following relation:
\begin{multline}
U_{\vec{k}+\vec{q'}}^{\dagger} \vec{q} \cdot \vec{\sigma} U_{\vec{k}+\vec{q'}} 
\\
= q 
\begin{pmatrix} \cos(\theta_{\vec{q}} - \theta_{\vec{k}+\vec{q'}}) & i \sin(\theta_{\vec{q}} - \theta_{\vec{k}+\vec{q'}}) \\ 
-i \sin(\theta_{\vec{q}} - \theta_{\vec{k}+\vec{q'}}) & -\cos(\theta_{\vec{q}} - \theta_{\vec{k}+\vec{q'}}) \end{pmatrix},
\end{multline}
Once we have the cut-off we can shift the sums:
\begin{equation}
\la 0|[[n_{\vec{q}}, H],n_{\vec{q}}^{\dagger}]|0 \ra = -v_F \Bigl[ \sum_{\vec{k} \in I } \cos ( \theta_{\vec{k}}) 
- \sum_{\vec{k} \in II } \cos ( \theta_{\vec{k}} ) \Bigr],
\end{equation}
where only the contribution from the lower band survives, and the regions of summation are shown in Fig.~\ref{figure_intregions}.
For a large momentum space cut-off $\Lambda = \Lambda_E / (2 v_F)$, the calculation of the difference can be carried out easily. Both regions give the same contribution but with opposite sign due to the cosine term. By going to the continuum limit of the sum we find:
\begin{equation}
\la 0|[[n_{\vec{q}},H],n_{\vec{q}}^{\dagger}]|0 \ra = \frac{q^2 \Lambda_E}{8\pi}.
\end{equation}
Reinstating the degeneracy of spin and valley gives rise to an extra multiplicative factor $g_v g_s = 4$. The final result for the f-sum rule is Eq.~\eqref{grapheneSumRule}:
\begin{equation}
\int_0^{\Lambda_E} d\omega \omega \Im \chi(\vec{q},\omega) = - \frac{g_S g_v q^2 \Lambda_E}{16} .
\nonumber
\end{equation}

\subsection{Doped graphene}

The effect of doping the graphene sheet translates in a nonzero chemical potential. The ground state no longer has a particle-hole symmetry. Therefore, the modifications to the f-sum rule calculation of the last section are: (1) The Hamiltonian must be replaced by $H-\mu N$, with $\mu$ being the chemical potential. (2) The ground state has a contribution from electrons of the upper band, for positive chemical potential, or from holes in the lower band, for negative chemical potential.

First we discuss (1). We have $N = n_{\vec{0}}$, which leads to $[n_q, N] = 0$, thus giving no
new contribution. Therefore, at the level of commutators the result is again Eq.~\eqref{grapheneSumRule}. 
From (2) we could expect a contribution from the electron (holes) in the ground state, above (below) the Dirac point. However, this contribution is no longer unbounded (the operators are only nonzero under (over) the Fermi momentum), and now we can shift the operators:
\begin{multline}
\sum_{\vec{k}} \cos(\theta_{\vec{q}} - \theta_{\vec{k}+\vec{q}}) 
\la 0|c_{\vec{k}+\vec{q}}^{\dagger} c_{\vec{k}+\vec{q}}|0 \ra 
\\
- \sum_{\vec{k}} \cos(\theta_{\vec{q}} - \theta_{\vec{k}}) 
\la 0|c_{\vec{k}}^{\dagger} c_{\vec{k}}|0 \ra = 0  .
\end{multline} 
Therefore, there is no contribution coming from the ground state evaluation of the commutators, and the result is the same than the undoped one, Eq.~\eqref{grapheneSumRule}.

\,
\,
\,

\subsection{Massive Dirac electrons} 

The case of massive electrons requires more attention. The Dirac Hamiltonian in this case reads:
\begin{equation}
H_0 = \sum_{\vec{k}} \Psi_{\vec{k}}^{\dagger} 
\begin{pmatrix} m & v_F (k_x - i k_y)\\ v_F(k_x + i k_y) & -m \end{pmatrix} \Psi_{\vec{k}} ,
\label{mass_ham}
\end{equation}
which is the same as Eq.~\eqref{Hamiltonian} with the additional term $m \sum_k \Psi_{\vec{k}} \sigma_z \Psi_{\vec{k}}$. Again, modifications to the f-sum rule could arise from the new term in the commutator or from the final ground state evaluation. The first contribution can be readily seen to be zero: $[n_q, H_m]=0$, due to the cancellation of the sublattice contributions independently. 
The second modification is more subtle, as requires the diagonalization of Hamiltonian in Eq.~\eqref{mass_ham}. The result gives an hyperbolic dispersion relation of the form $E_k = \sqrt{v_F^2 k^2 + m^2}$. Applying it to the commutators which give the f-sum rule (following the same lines than above) we get:
\begin{multline}
\la 0|[[n_{\vec{q}}, H],n_{\vec{q}}^{\dagger}]|0 \ra = \\
=  - v_F \Bigl[ \sum_{\vec{k} \in I}  \frac{v_F k}{\sqrt{(v_F k)^2+m^2}}  \cos (\theta_{\vec{k}}) 
\\
- \sum_{\vec{k} \in II}  \frac{v_F k}{\sqrt{(v_F k)^2+m^2}}  \cos(\theta_{\vec{k}}) \Bigr].
\end{multline}
The result again involves the subtraction of two regions, which is nonzero only for momenta close to the cut-off, as we showed in Fig.~\ref{figure_intregions}. If $m<<\Lambda_E$ we can expand the prefactor to leading order in $k$, recovering the massless result. This is to be expected since the mass term is irrelevant in this region. The result for the massive case turns out to be the same than the massless one of Eq.~\eqref{grapheneSumRule}, so the existence of a gap does not change the nature of the sum rule. At this point we remark that the particular form of the f-sum rule for Dirac electrons comes mainly from the asymptotic linear spectrum and the existence of an unbounded spectrum.

\bibliography{spectral}
\end{document}